# Optimization of Service Addition in Multilevel Index Model for Edge Computing

Jiayan Gu, University of Leicester, Leicester, UK
Yan Wu, University of Jiangsu, Jiangsu, China
Ashiq Anjum, University of Leicester, Leicester, UK
John Panneerselvam, University of Leicester, Leicester, UK
Yao Lu, University of Leicester, Leicester, UK
Bo Yuan, University of Derby, Derby, UK

Corresponding authors: Ashiq Anjum, aa1180@leicester.ac.uk; Yan Wu, wuyan04418@ujs.edu.cn

*SUMMARY*
**With the development of Edge Computing and Artificial Intelligence (AI) technologies, edge devices are witnessed to generate data at unprecedented volume. The Edge Intelligence (EI) has led to the emergence of edge devices in various application domains. The EI can provide efficient services to delay-sensitive applications, where the edge devices are deployed as edge nodes to host the majority of execution, which can effectively manage services and improve service discovery efficiency. The multilevel index model is a well-known model used for indexing service, such a model is being introduced and optimized in the edge environments to efficiently services discovery whilst managing large volumes of data. However, effectively updating the multilevel index model by adding new services timely and precisely in the dynamic Edge Computing environments is still a challenge. Addressing this issue, this paper proposes a designated key selection method to improve the efficiency of adding services in the multilevel index models. Our experimental results show that in the partial index and the full index of multilevel index model, our method reduces the service addition time by around 84% and 76%, respectively when compared with the original key selection method and by around 78% and 66%, respectively when compared with the random selection method. Our proposed method significantly improves the service addition efficiency in the multilevel index model, when compared with existing state-of-the-art key selection methods, without compromising the service retrieval stability to any notable level.**

*KEYWORDS*
**Edge computing, Service computing, Big data management, Multilevel index**

1. INTRODUCTION

The rapid development of the concept of Artificial Intelligence (AI) and Big Data has transformed the way services are being offered over the Internet. The unprecedented growth in the number of real-time data flow applications has led to an increase in the deployments of service architectures. Numerous information services are stored in large-scale service repositories and are made available to customers via Cloud-based service models.[1] The flexible deployment and low coupling characteristics of services repositories make services easier to manage and maintain. In addition, a lot of intelligent applications are being deployed at the edge nodes using the container technology and service architectures.[2] Upon a service request from customers, a service model should be able to efficiently and accurately discover the requested information services with reduced network latency. Whilst the services stored in large-scale service repositories increase the scope of service availability to satisfy the various needs of customers, this also adds burden on the network since such service models involve data communication between the clients and the Cloud servers, thereby resulting in issues such as network bottleneck, additional latency, and bandwidth unavailability etc.[3] For instance, applications such as driverless vehicles relying on Cloud servers are extremely latency sensitive, usually in the ranges of milliseconds. Undesirable latencies in the network can cause a significant impact on the service quality and could even lead to the failure of such applications. Recently, edge computing is emerging as a potential solution to address the latency issues of the traditional Cloud-based service models.[4] Edge computing can host considerable proportions of the process execution at the edge of the network than uploading all the data to the Cloud servers.[5] Such a strategy can significantly help to reduce the pressure on network bandwidth and also the power consumption of the data centres.[6] Furthermore, hosting the data processing near the source where possible can help reduce the round-trip time, which greatly reduces the delay in the entire processing architecture and enhances the service response capability.[7] This recent evolution of edge computing has accelerated the development of a large number of edge services.

Edge computing has attracted data intensive and latency sensitive applications and exploits the capabilities of the edge devices to collect data from the sensor nodes and process the collected data in the edge network where possible. An increasing number of edge devices alongside the deployments of millions of industrial sensors in the physical and artificial environment is generating large volumes of complex real-time streaming data, leading to ever-changing ways of service discovery and composition, and the types of services invoked by users continue to expand.[8,9] Such an unprecedented data generation is posing several challenges in acquiring and processing sensor generated data.[10] The real-time processing requirements of big data applications are making the traditional Cloud-based services models unsuitable for latency sensitive applications. Furthermore, real-time invocation of services is quite challenging in the traditional Cloud service models for applications involving intensive data processing. Edge computing can be a potential alternative to host most of the services processing near the source of data generation, which requires suitable strategies at the software-tier to exploit the capabilities of intelligent devices that are deployed in the edge network. In most edge applications, an orchestration of the edge devices is required to enhance the computational capabilities of the processing cluster in order to accommodate intensive executions. For instance, edge networks in a smart home scenario are used to control the operations services of the household devices such as lights, robots, automated guided vehicle (AGV)[11] and video image recognition by the real-time services based on the Internet of Thing are increasingly being deployed. Pang et al. said that Edge computing is being promoted as a strategy to achieve scalable and highly available Web services.[12,13] It pushes data processing from cloud data centres out to proxy servers at the "edge" of the network.[14,15] Web services could turn into a service layer to the edge, where application logic would be utilized to gather applications and data on the fly from multiple origin servers for access by nearby end-users. However, efficient service composition and discovery are becoming a challenging problem in service applications, since such an application usually comprises a large number of different services with complex interactions.[16] It is worthy of note that the mentioned services are usually distributed in the edge devices, and identifying the required services is an important task so that the processing cluster can be formed with the right capacity. To this end, strategies for invoking and orchestrating edge devices with improved service discovery and composition efficiency are being widely researched in recent years. Some researchers have postulated the use of various service index structures to reduce the service composition time, such as service net,[17] inverted index,[18] T-tree,[19] and multilevel index model.[20,21] Especially, the multilevel index model proposed by Y. Wu et al. is a large-scale service storage structure based on equivalence relation and quotient theory, which can effectively reduce the service retrieval and composition time. Researchers have postulated various propositions on this multilevel index model. Z. Xu and Y. Wu used the chord protocol to the multilevel index model to simulate the distributed architecture.[22] D. Miao et al. proposed the index model on the fog layer to improve the service discovery efficiency.[23] They insert a newly generated service into the matched fog node with the DM-index model to retrieve whether the service exists in the fog node. W. Xu et al. and Y. Wu et al. further studied the performance of the multilevel index model.[24,25] When a service is added into the multilevel index, an input parameter is selected as a key for the service. They argued that the service retrieval process is not related to the service keys. W. Kuang et al. investigated these propositions and proposed a random key selection method to improve the stability of the service retrieval process and to reduce the service addition time significantly.[26] Our work proposed in this paper extended the multilevel index model in the edge environment, which can effectively manage big data and proposes a designated key selection method, with the motivation of narrowing down the search space of service retrieval process and to reduce the time overheads without affecting the stability of the service retrieval process. The characteristics of the proposed key selection method are adaptive to the dynamic changes of services commonly witnessed in the edge environment.

The main contributions of this work are summarised as follows:

1. An optimised multilevel index model has been proposed for the efficient management and discovery of services in dynamic edge computing environments.

2. A designated key selection method based on the service addition algorithm is proposed, which can reduce the service addition time without compromising the service retrieval efficiency and stability.

3. The time complexity of the proposed service addition method is analysed. Furthermore, the proposed service addition process is mathematically modelled to analyse its efficiency.

4. A simulation platform of the multilevel index model is implemented with the integration of the designated key selection method and corresponding service addition operation. Experiments validated the efficiency of our proposed service addition algorithm.

The rest of this article is organized as follows: Section II reviews the latest related work on the effective service discovery and composition of the Internet of Things in the edge environment. Section III introduces the multilevel index model in the edge environment. Section IV introduces the designated key selection algorithm proposed in this paper and analyses the time complexity of the proposed service addition algorithm. Section V presents and discusses the validity of the experimental results. Section VI summarizes the full paper and highlights future research directions.

2. RELATED WORK

In Cisco's 2018-2023 Annual Internet Report,[27] it is clear that by 2023, the total number of Internet users will reach 5.3 billion (66% of the global population), up from 3.9 billion (51% of the global population) in 2018. By 2023, the number of network devices will reach 29.3 billion, up from 18.4 billion in 2018. Meanwhile, 299.1 billion mobile applications will be downloaded globally. Given the growth of data generated in the recent past, the concept of processing services data at the edge networks is of growing importance. With the concept of hosting the data processing at the edge, the amount of data transmitted to the cloud can be significantly reduced, which further reflects in the reduction of bandwidth consumed of network resources. This ultimately reduces the operational costs of the entire application. Given such attractive features of edge computing, numerous researches have been initiated in the recent past focusing on developing efficient services at the edge networks. Concepts such as cloudlet,[28] big data[29] and services[30] have been introduced into the context of edge processing. T. Wang et al. projected that a large number of integrated sensors to collect multi-feature data will be added to the sensor cloud system. In addition to the actual data execution, pre-processing of the data such as data cleansing methodologies for industrial standards have also been proposed in the edge networks, based on mobile edge nodes. Such strategies help not only to ensure reliability but also to verify data integrity, alongside reducing the bandwidth and energy consumption of the industrial supply chain process.[10] Industrial instruments may involve a large number of difficult-to-control sensors, which not only overloads the network capacity, but also generates a lot of garbage data that are not actually required for analytics. Thus, it is important to effectively retrieve the data that contributes towards application analytics before their transmission. Besides, the advent of huge amounts of data also means the growth of a large number of services. The discovery, composition, and invocation of service under edge environments is gaining the attention of many researchers and requires more strategic approaches to enhance the processing efficiency of edge networks. It is obvious that the ever-increasing amount of services requires a highly scalable storage solution, and A. Miranda et al.[31] proposed a random slicing strategy that combines lessons learned from table-based, rule-based, and pseudo-random hashing strategies to provide a simple and effective strategy for processing exascale data. However, with the expansion of data volume, more and more different types of services are created. A service index model designed according to service characteristics needs to be proposed to manage and store services more efficiently, thereby effectively improving service discovery and service composition.

Z. Xu et al. and D. Miao et al. proposed the multilevel index model in distributed architectures, which can effectively reduce the time of service retrieval.[22,23] This method still used the traditional key selection method for providing keys for every service. They further attempted to prove that the original key selection method is the most efficient for service retrieval in the multilevel index model under an assumption of the same invocation probabilities for all the services, which may not be true in a large-scale distributed service repository. Moreover, W. Xu et al. and Y. Wu et al. uncovered that in the multilevel index model, the original key selection method characterises the assumption of the time complexity of comparing two different elements being equal for two different sets, again this may not be true all the time.[24,25] W. Kuang et al. postulated that although the key selection methods have considerable impacts on the service retrieval stability and performance, their impacts on average retrieval time is negligible.[26] Based on such findings, they proposed a random key selection method which reduces the service addition time and further improves the stability of the service retrieval process. However, the randomness of the random key selection method is obvious, and whether it can be the optimal key selection method is not persuasive. Based on the aforementioned discussions, we propose a convincing method for designated key selection verified by the experiments, which can significantly improve the service addition efficiency without compromising the service retrieval efficiency and stability.

## 3. MULTILEVEL INDEX MODEL FOR EDGE COMPUTING

Edge computing characterises a distributed processing architecture, where possible proportions of data processing are hosted to release the pressure on the networks and at the back-end datacentres. In the edge computing model, caching and retrieving data in the edge servers can be more efficient due to the reduction in the bandwidth requirements and network latency. In general, each edge node in the edge environment may have a service repository. Sensors deployed in the field-of-view are usually responsible for collecting, processing, and uploading data to the edge server. The edge servers, when receiving the data, execute and store the data, and facilitate the required services to the devices in its surrounding vicinity. Such edge devices are often located closer to the users and can pre-process the collected service information to meet the real-time requirements, as shown in Figure 1.

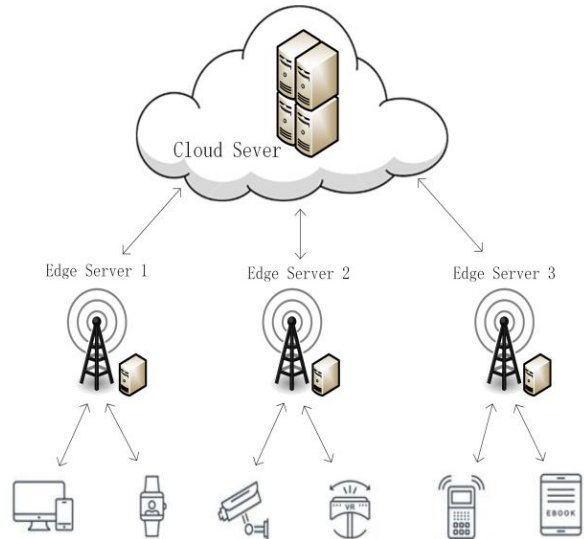

Figure 1. Retrieving data in the edge computing environment.

Edge servers can use the multilevel index model, shown in Figure 2, to facilitate edge service operation and management. Since the service retrieval and addition operations play an effective role in the multilevel index model, the utilisation of the multilevel index model in edges servers is very suitable for dynamic changes witnessed in the edge services.

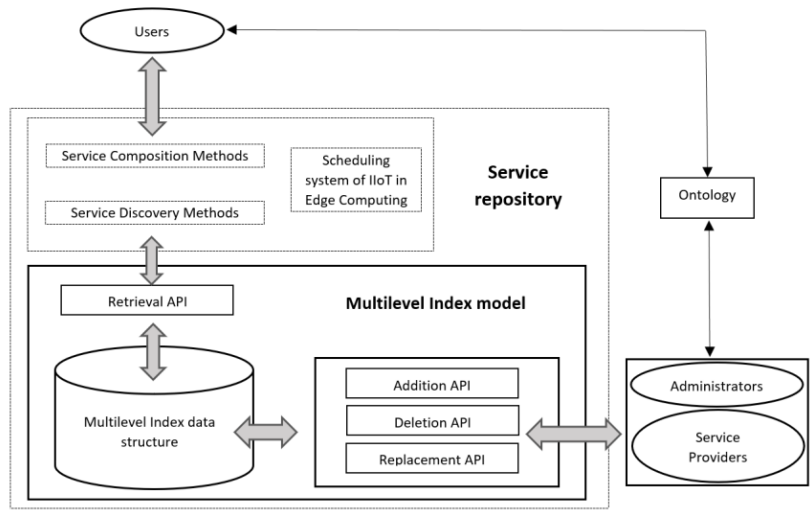

Figure 2. The framework of the multilevel index model.

3.1 Basic definition of service

In a service repository, each service has a set of input parameters and a set of output parameters, which can be defined as: $s= (•s, s•)$, where $•s$ indicates the set of input parameters, and $s•$ indicates the set of output parameters.[32] According to P. Bartalos and M. Bielikova,[33] suppose that an $s= (•s, s•, Pre, Post)$ is a Web Service, where *Pre* and *Post* represent the conditions that can be invoked for the corresponding Web Service (pre-conditions) and the conditions that are retained after execution (post-conditions), respectively. However, services not only include the input parameters and output parameters, but also the important QoS to represent the quality of service retrieval and BC to represent a set of behaviour constraints.

In order to comply with the above service definition, the service retrieval operation proposed by Wu et al.[20,21] is revised as follows:

**Definition 1.** A service is represented as a tuple $s = \{•s, s•, O\}$, where $•s$ is the set of input parameters, and $s•$ is the set of output parameters, and $O$ is a set of service attributes, e.g., QoS.

**Definition 2.** Service retrieval is represented as a tuple $Re(A, S) = \{s | •s \subseteq A \wedge s \in S\}$, where $A$ is on behalf of a given parameter set and $S$ represents a service set. Service retrieval can be defined as the process of finding all the services whose input parameters are contained in $A$ from the service set $S$, which means all the services that can be invoked under $A$.

3.2 Multilevel Index Model

The multilevel index model is built based on the equivalence relation, with the motivation of reducing the redundancies resulting in the service retrieval process.[20,21] The multilevel index model is divided into four levels, described as follows:

The First Level Index ($L_1I$): This is an index between a service s and a similar class $C_s$ if $s \in C_s$. All similar services with the same input and output parameters are clustered into a class, which constructs the First Level Index. L1I reduces the redundancies introduced by these services with the same input and output parameters.

The Second Level Index ($L_2I$): This is an index between a similar class $C_s$ and an input-similar class $C_{is}$ if $C_s \in C_{is}$. All similar services with the same input parameters are clustered into a class, which constructs the Second Level Index. $L_2I$ reduces the redundancies introduced by these services with the same input parameters.

The Third Level Index ($L_3I$): This is an index between an input-similar class $C_{is}$ and a key class $C_k$ if $C_{is} \in C_k$. The third level index cannot be used by the service retrieval operation individually, where the co-operation of the fourth level index is required, so that the service retrieval efficiency can be improved.

The Fourth Level Index ($L_4I$): This is an index between a key class $C_k$ and a parameter $k$.

Three different deployment schemas for the 4 level indices have been proposed by Y. Wu et al.[21] They are the primary index model ($L_3I$-$L_4I$), partial index model ($L_2I$-$L_4I$) and full multi-level index model ($L_1I$-$L_4I$). the structures of these models are illustrated in Figure 3 to Figure 5.

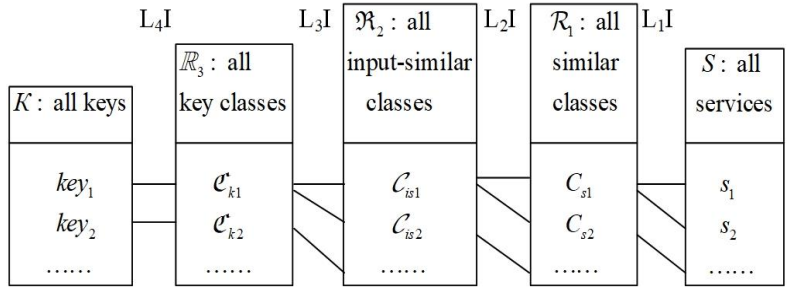

Figure 3. Full index model deployed by $L_1I$-$L_4I$.

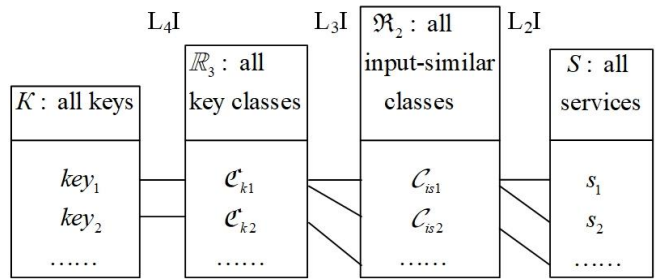

Figure 4. Partial index model deployed by $L_2I$-$L_4I$.

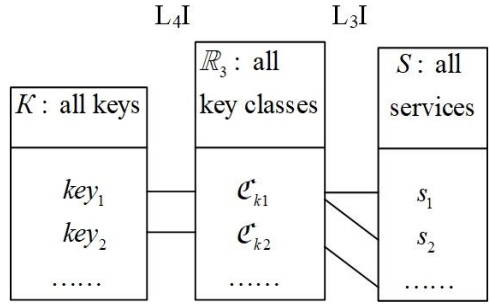

Figure 5. Primary index model deployed by L$_3$I-L$_4$I.

3.3 Different Key Selection Methods

The service information addition operation in the service repository uses the addition algorithm to take one of the input parameters of the selected service as its key and insert the service information into the multilevel index model. Currently, there are different key selection methods in the multilevel index model for adding services. The traditional key selection method[20] proposed in selects keys in such a way that $|\mathcal{C}_k|$ is maintained as close as possible to $\sqrt{|S|}$ in the primary index or $\sqrt{|\mathfrak{R}_2|}$ in the partial/full indices, respectively.

Y. Wu et al. introduced the expected value of traversed service count as an indicator to compare the retrieval performances of different index models.[25] They analysed the expected values of the traversed service counts of the primary, partial and full index models, and their expressions denoted as $E_{pr}$, $E_{pt}$ and $E_{fl}$, respectively, as below.

$$E_{pr} = \frac{r}{|P|} \times |S| \qquad (1)$$

$$E_{pt} = \frac{r}{|P|} \times |\mathfrak{R}_2| \qquad (2)$$

$$E_{fl} = \frac{r}{|P|} \times |\mathfrak{R}_2| \qquad (3)$$

where, $r$ denotes the average count of parameters of each retrieval request, $|P|$ denotes the parameter count of all service inputs, and $|S|$ and $|\mathfrak{R}_2|$ denotes the size of the service set and the input-similar set.

Based on the above studies, W. Kuang et al. identified that the expected values do not relate to the size of the key set, i.e., $|K|$. Three effective methods of key selection have been introduced in their work, which are the maximum key selection, the minimum key selection and the random key selection.[26] The principle of the maximum key selection method is to make the key set size as large as possible. On the contrary, the principle of the minimum key selection method is to keep the key set size as small as possible. However, the random key selection method is different from the other two, which randomly chooses an input parameter of a new service as its key.

The experimental analysis presented in the literature postulates that the key selection methods do not affect the service retrieval performance but exerts considerable impacts on service retrieval stability. If the response times of the two retrieval requests differ greatly, it means that the retrieval process is not stable enough. The random key selection method is recognised as the most optimal key selection method for achieving stability, among the state-of-the-art methodologies.

An unexpected benefit is that the random key selection method improves the service addition performance. But the reason behind this has not been explained by W. Kuang et al.[26] We further explore the reason behind this characteristic of the random key selection method and propose a better key selection method in the following section.

## 4. OPTIMIZATION OF SERVICE ADDITION

In this part, without affecting the efficiency and stability of service retrieval, a designated key selection method of service addition algorithm is proposed and integrated into the multilevel index model to reduce the time of service addition. In addition, the time complexity of the proposed service addition algorithm is analysed, and the mathematical model of the process of service addition is further established to analyse the efficiency of service addition.

4.1 Designated Key Selection Method

Service addition algorithms for a primary index model can be described as follows.

---

***Algorithm*** 1: Service addition for primary index
Input: a service, a primary index;
Output: a primary index added with the service;
1. use a key selection method to select a key for the service;
2. add the service into the primary index according to the key.

---

Service addition algorithms for a partial/full index model can be described as follows.

---

***Algorithm*** 2: Service addition for partial/full index
Input: a service, a partial/full index;
Output: a partial/full index added with the service;
1. for each input parameter of the service
2. {if the input parameter is a key
3. {determine in the key class mapped by the key whether existing an input-similar class such that its input set equal to the service input set;
4. if existing
5. {add the service into the input-similar class;
6. return;}}}
7. use a key selection method to select a key for the service;
8. add the service into a new input-similar class;

---

Line 1 in ***Algorithm*** 1 and line 7 in ***Algorithm*** 2 invokes a key selection method respectively. Among the four key selection methods tested, the random key selection[26] is identified to be the simplest method that randomly selects an input parameter of service as its key. Therefore, it provides the most optimum performance for service addition among the four methods.

In ***Algorithm*** 2, line 1-6 determines whether an existing input-similar class has its input set equal to the service input set, which can guaranty that the set of the input-similar class is a quotient set of services in the partial index model or similar classes in the full index model. This process is similar to service retrieval, but it is time consuming. Now, we propose a new key selection to narrow down the search space when adding a new service into a partial or full index. Since the random key selection method exhibits the highest stability and performance, we use a pseudorandom method to simulate the random key selection method, whilst maintaining the correctness of the quotient set. The proposed designated key selection method is shown as follows.

---

***Algorithm*** 3: Designated key selection method
Input: a service;
Output: key of the service;
1. *sum*=0;

   2. For each input parameter *a* of the service
   3.    {*sum=sum+a.id*;}
   4. *i=sum* mod *c*; // (*c* denotes the count of the input parameter of the service.)
   5. return the $i^{th}$ input parameter as the key of the service.

---

Based on ***Algorithm*** 3, ***Algorithm*** 2 can be simplified as follows.

---

***Algorithm*** 4: Service addition for a partial/full index
Input: a service, a partial/full index;
Output: a partial/full index added with the service;
   1. use Algorithm 3 to select a key for the service;
   2. determine in the key class mapped by the key, whether an existing input-similar class has its input set equal to the service input set;
   3. if existing
   4.    {add the service into the input-similar class;
   5.     return;}
   6. else
   7.    add the service into a new input-similar class;

---

The correctness of ***Algorithm*** 4 is obvious. Suppose that, service $s_1$ and service $s_2$ have the same input parameters, and then their sums computed by line 2 and 3 in ***Algorithm*** 3 are the same, and the $i^{th}$ values computed by line 4 are the same. Since the input parameters of the two services are ordered by the same rules, their keys are the same, so that they are added into the same key class. Line 2 in ***Algorithm*** 4 determine whether an input-similar class exists in the key class such that its input set is equal to the service input set. If so, the key class is added into the same input-similar class. If not, the first service is added into a new input-similar class with the same input parameter set. The second service is added into the same input-similar class since their keys and input parameters are the same. Therefore, ***Algorithm*** 4 is correct.

In ***Algorithm*** 2, lines 1-6 determine whether every input parameter contains a key in the key set $K$. If keys exist, then the input-similar classes contained in the corresponding key class are checked to identify whether their input parameter sets are equal to the service's input parameter set. Many such keys are checked in this process. Whilst in ***Algorithm*** 4, lines 1-5 determine only one key class. Therefore, ***Algorithm*** 4 reduces the time involved in this key checking process in comparison with ***Algorithm*** 2.

According to the above analysis, it is evident that the proposed designated key selection method can improve the performance of the service addition process for partial and full index models. However, the designated key selection method cannot improve the service addition operation for the primary index model since this index does not contain any input similar class, where the random key selection method performs better.

4.2 Expectations of service addition for three multilevel indices

Y. Wu et al. used the expected value of the traversed service count as an indicator to compare the retrieval performances.[25] Herein, the expected value of the compared parameters of service is used as an indicator in this paper to compare the addition performances. $A_{pr}$, $A_{pt}$ and $A_{fl}$ denotes the expected values of the primary, partial and full index models, respectively.

For the primary index model, $A_{pr}^R$ and $A_{pr}^D$ denote the expectations of service addition using the random key selection method and the designated key selection method, respectively. When a key of a new service is selected, the addition operation needs to retrieve the key from the key set $K$ and then the relevant key class where the service should be added is determined. Since all the keys are stored in a binary tree structure, $\log_2|K|$ number of keys are required to be compared to find the key class. Expectations of service addition for the primary index using the random key selection method and the designated key selection method are equal. Their formulas are presented as follows.

$$A_{pr}^R = \log_2 |K| \qquad (4)$$

$$A_{pr}^D = \log_2 |K| \qquad (5)$$

For the partial index, expectations of service addition using the random key selection method and the designated key selection method are denoted as $A_{pt}^R$ and $A_{pt}^D$, respectively. Their formulas are presented as follows.

$$A_{pt}^R = n \times \log_2 |K| + \frac{|K|}{|P|} \times n \times \frac{|\Re_2|}{|K|} \times n \qquad (6)$$

$$A_{pt}^D = \log_2 |K| + \frac{|K|}{|P|} \times 1 \times \frac{|\Re_2|}{|K|} \times n \qquad (7)$$

where, $n$ denotes the average count of the service input parameters.

For the partial index using a random key selection method, i.e., $A_{pt}^R$, according to **Algorithm** 2, when a new service is added, the addition operation checks each input parameters to determine whether it is a key or not, and thus $n \times \log_2|K|$ parameters are compared. $|K|/|P|$ means that all keys are distributed on $|P|$. $|K| \times n/|P|$ denotes the number of input parameters that can be used as keys for a service. $|\Re_2|/|K|$ denotes the number of input-similar class those are linked by a key. $(|K|/|P|) \times n \times (|\Re_2|/|K|)$ denotes the total number of input-similar classes compared. Since each input-similar class contains $n$ input parameters, $(|K|/|P|) \times n \times (|\Re_2|/|K|) \times n$ input parameters are compared. Therefore, $A_{pt}^R$ parameters are compared in the entire process. Using the designated key selection method, i.e., $A_{pt}^D$, according to **Algorithm** 4, the key of the service has been determined, where just one key and one key class are checked. Therefore, $A_{pt}^D = A_{pt}^R/n$.

For the full index, expectations of the service addition process using the random key selection method and the designated key selection method are denoted as $A_{fl}^R$ and $A_{fl}^D$, respectively. Their formulas are presented as follows.

$$A_{fl}^R = n \times \log_2 |K| + \frac{|K|}{|P|} \times n \times \frac{|\Re_2|}{|K|} \times n + \frac{|\Re_1|}{|\Re_2|} \times m \qquad (8)$$

$$A_{fl}^D = \log_2 |K| + \frac{|K|}{|P|} \times 1 \times \frac{|\Re_2|}{|K|} \times n + \frac{|\Re_1|}{|\Re_2|} \times m \qquad (9)$$

where, $m$ denotes the average count of service output parameters.

When compared to $A_{pt}^R$, the addition operation in the full index needs to select a similar service in order to add a new service. $|\Re_1|/|\Re_2|$ denotes the number of similar classes linked by an input-similar class. $(|\Re_1|/|\Re_2|) \times m$ denotes the number of parameters compared whilst finding a similar class to add the service. Therefore, $A_{pt}^R = A_{fl}^R + (|\Re_1|/|\Re_2|) \times m$. Since keys do not affect L₂I, $A_{pt}^D = A_{fl}^D + (|\Re_1|/|\Re_2|) \times m$.

From the above analysis, it can be stated that the designated key selection method does not affect the addition operation in a primary index but can significantly reduce the service addition time for partial and full indices. We prove this claim through experimental evaluations in the next section.

5. EXPERIMENTS AND ANALYSIS

The simulation experiments conducted on a personal laptop which has an Inter Core i7-3540M CPU with 3.00 GHz and a RAM with 8.00 GB and the experimental platform is developed in Microsoft Visual Studio 2017 using Visual C# object-oriented development technology. The advantage is that the coupling of each component is low, which can be modified, upgraded or replaced separately from the application layer module to the structural layer module. Each of the synthetic datasets contains 20,000 services and 1000 parameter sets, i.e., $|S|=20,000$ and $|P|=1000$, respectively. Each service has 10 input and 10 output parameters, i.e., $n=10$ and $m=10$. Each retrieval request contains 32 parameters. A total of 10 datasets have been generated in this experiment and every dataset contains 100 retrieval requests except the datasets that are used to test the stability, which contains 1 retrieval request.

Our proposed designated key selection method not only preserves the stability of the service retrieval process, but also reduces the service addition time better than the original key selection and random key selection. In order to demonstrate this, three different types of multilevel index model are generated for the service addition operations using the original key selection method, the random key selection method, and the designated key selection method, respectively.

Figure 6 illustrates the retrieval time performance of the three key selection methods in the primary index model. It can be observed that the three key selection methods exhibit similar performance, implying that the key selection method does not affect the retrieval efficiency to any notable extent.

Figure 7 depicts the traversal service count of 1000 requests in the primary index model. The service retrieval performance of the original key selection method is not stable, while the service stability performance of the random key selection and our designated key selection method is very similar. From[26], we know that the random key selection method exhibits the most stable service retrieval performance in the state-of-the-art so far. Therefore, the designated key selection method outperforms the existing key selection methods in terms of service stability performance.

The partial indices and full indices also present similar results, but the results are not discussed in detail due to space constraints.

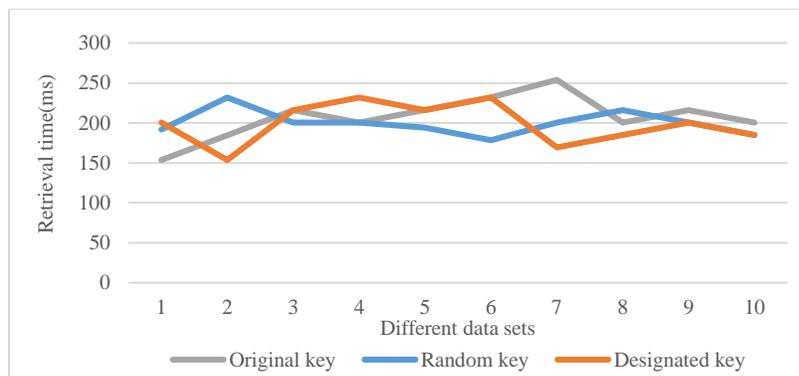

Figure 6. Service retrieval time for 10 test sets in primary index models generated by original, random and designated key selection methods.

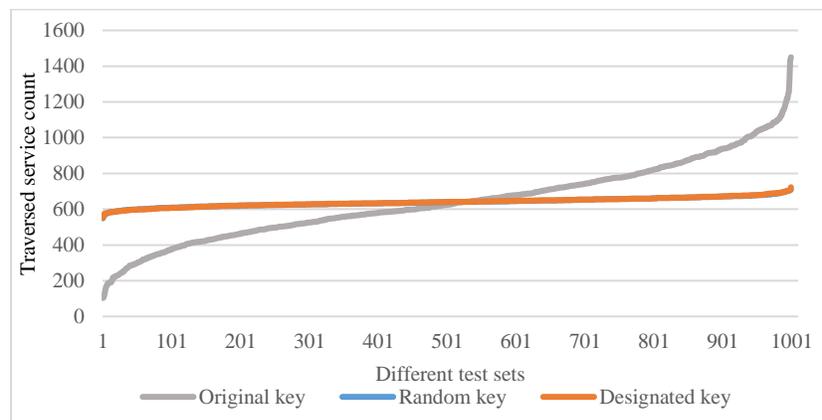

Figure 7. Traversed service counts for 1000 retrieval requests in primary index models generated by original, random and designated key selection methods.

Figure 8 presents the service addition time for 10 test datasets in the primary index model with the original, the random and the designated key selection methods respectively. The designated key selection method should undergo a few computations in order to determine the key, but the random key selection method does not involve any computation. This impact of the additional computation incurred in the designated method is evident in Figure 8, where the designated method characterises a slightly longer service addition time than the random key selection method, but its service addition time is significantly shorter than that of the original key selection method without any optimisation method.

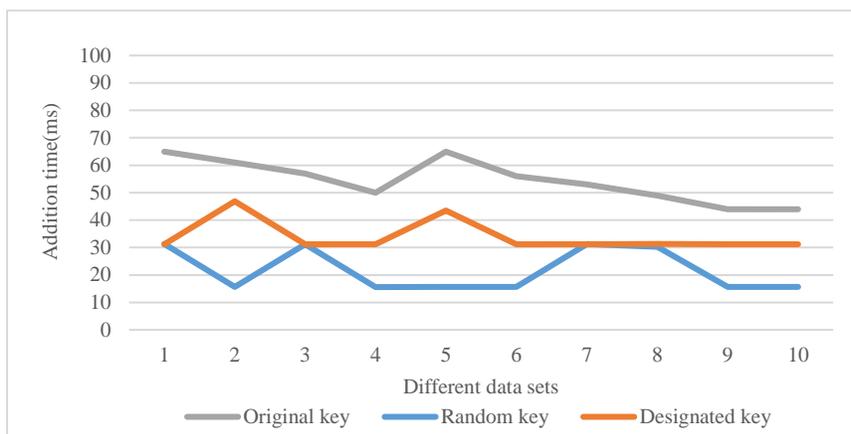

Figure 8. Service addition time for 10 test sets in primary index models generated by original, random and designated key selection methods.

Figure 9 compares the service addition time of the three key selection methods in the partial index model. It is obvious that the service addition time of our proposed designated key selection method is far less than that of the random key selection method, which conforms with our analysis presented in section IV. Our proposed designated key selection method improves the efficiency of the addition operation by nearly 78% than that of the random key selection method and nearly 84% than that of the original key selection method, respectively.

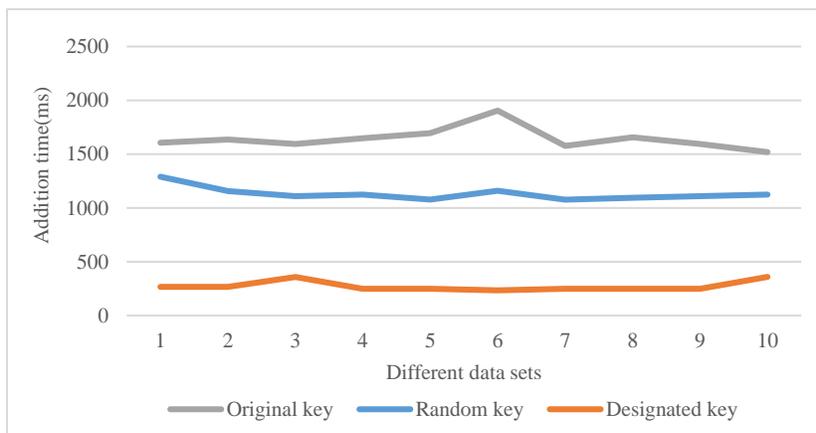

Figure 9. Service addition time for 10 test sets in partial index models generated by original, random and designated key selection methods.

Figure 10 compares the service addition time of the designated key selection method, the random key selection method and the original key selection method in the full index models. The experimental results are very similar to the results presented in Figure 9, and also verify the correctness of our analysis presented in section IV. The performance of service addition operation based on the designated key selection method is much more efficient than that of using the original key and random key selection methods in the full index model. The structure of the full index model is more complex than the partial index model. Since the service addition operation needs to consider the average number of output parameters of services in the full index model, the overall service addition time in the full index model is slightly higher than the service addition time in the partial index model. However, our proposed designated key selection method outperforms both the random key selection and the original key selection methods by 66% and 76%, respectively.

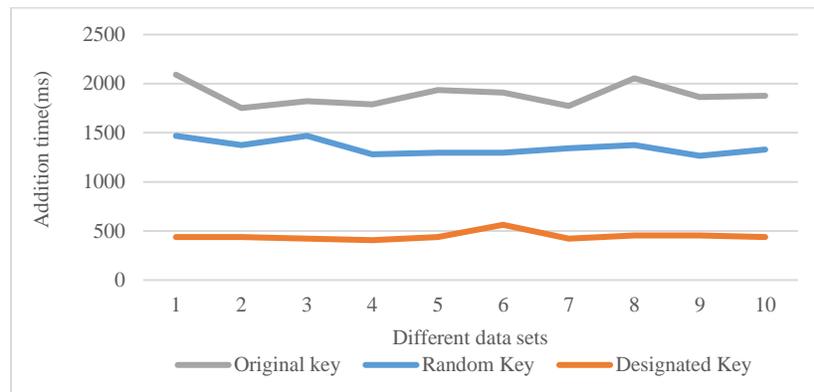

Figure 10. Service addition time for 10 test sets in full index models generated by original, random and designated key selection methods.

From the above experimental results, we can conclude that the designated key selection method proposed in this paper improves the service addition efficiency significantly in partial and full indices without any loss of service retrieval efficiency and stability.

6. CONCLUSION

This paper proposed a novel service addition algorithm with the development of a designated key selection method to improve the service addition efficiency of the multilevel index model in edge computing environments. Although the efficiency of the proposed designated key selection method based on the service addition algorithm is performing similar to that of the original key selection method and the random key selection method in the primary index model, our method significantly reduces the service addition time by around 84% and 76% in partial and full index models respectively than that of the original key selection methods, and around 78% and 66% in partial and full index models, respectively, than that of the random selection methods. The experimental results also conform to our theoretical verification. In addition, the proposed key selection method based on the service addition algorithm does not affect the service retrieval efficiency and stability, when compared to the random key selection method. As future work, we plan to integrate the proposed method of multilevel index model into the real-world edge devices, along with further optimising the ability of services management and improving the efficiency of services discovery in edge servers.

DATA AVAILABILITY STATEMENT
The data that support the findings of this study are available from the corresponding author upon reasonable request.

ORCID
Jiayan Gu https://orcid.org/0000-0001-9355-5395
Ashiq Anjum https://orcid.org/0000-0002-3378-1152
John Panneerselvam https://orcid.org/0000-0002-0332-1681
Bo Yuan https://orcid.org/0000-0001-8401-321X